\documentclass[useAMS, usenatbib, usegraphicx]{mn2e}
\usepackage{times}
\usepackage{natbib}
\usepackage{graphicx}
\usepackage{amsmath}
\newcommand{\str}{Str\"{o}mgren }
\newcommand{\ars}{R_{\rm s}}
\newcommand{\rhoH}{\rho_{\rm H}}
\newcommand{\rhoL}{\rho_{\rm L}}
\newcommand{\nH}{n_{\rm H, \infty}}

\newcommand{\nfive}{10^5~{\rm cm^{-3}}}

\newcommand{\none}{10~{\rm cm^{-3}}}

\newcommand{\hi}{\rm H~{\textsc i}}
\newcommand{\hii}{\rm H~{\textsc {ii}}}

\title[Rayleigh--Taylor instability of I-front around BHs]
{Rayleigh--Taylor instability of ionization front around black holes}
\author[K. Park, M. Ricotti, T. Di Matteo, and C.S. Reynolds]{KwangHo
Park$^{1}$\thanks{E-mail: kwanghop@andrew.cmu.edu}, Massimo
Ricotti$^{2,3}$, Tiziana Di Matteo$^{1}$, and Christopher S. Reynolds$^{2}$
\\ 
$^{1}$McWilliams Center, Carnegie Mellon University, Pittsburgh,
PA 15213, USA\\
$^{2}$Department of Astronomy, University of Maryland, College Park,
MD 20740, USA\\
$^{3}$Sorbonne Universit\'{e}s, Institut Lagrange de Paris (ILP), 98
bis Bouldevard Arago 75014 Paris, France} 

\begin{document}
\date{Accepted 2013 October 28. Received 2013 October 17; in original form 2013 August 22}

\pagerange{\pageref{firstpage}--\pageref{lastpage}} \pubyear{2013}

\maketitle

\label{firstpage}

\begin{abstract}
We examine the role of ionizing radiation emitted from black holes
(BHs) in suppressing the growth of the Rayleigh--Taylor instability
(RTI) across the ionization front (I-front) that forms when the gas
fuelling the BH is neutral. We use radiation-hydrodynamic simulations
to show that the RTI is suppressed for non-accelerating fronts on all
scales resolved in our simulations. A necessary condition for the stability
of the I-front is that the radius of the \str sphere is larger than the 
Bondi radius. When this condition is violated the I-front collapses
producing an accretion luminosity burst. Transient growth of the RTI
occurs only during the accretion burst when the effective acceleration
in the frame of reference of the I-front increases significantly due
to the rapid expansion of the \str sphere.
\end{abstract}


\section{Introduction}
Rayleigh--Taylor instability (RTI) develops across an interface
between two media with different densities with a presence of
background gravitational acceleration or when the density front is
accelerating \citep{Taylor:50,Chandrasekhar:61}. Gravity accelerates
high-density gas into the low-density medium, leading to an exponential
growth of the instability. In astrophysical circumstances,
special attention has been paid to the RTI coupled with radiation
field in star-forming \hii~regions or black hole (BH) systems
\citep*{Kahn:58, Krolik:77, KrumholzM:09, JacquetK:11, KuiperKBH:12, JiangDS:13}.
In particular, \citet*{Mizuta:05} studied the growth of RTI at
the ionization fronts (I-fronts) produced by UV photons from massive
star formation and showed that recombination in the ionized gas
suppresses the growth of RTI.

Similarly, ionizing UV and X-ray photons emitted by BHs embedded in a
neutral medium produce a {\it D}-type (dense) I-front that should be
unstable to RTI. The hydrogen ionizing photons heat the surrounding
gas that expands producing a hot, rarefied and ionized medium opposing
the collapse of the denser ambient medium subject to the BH
gravitational field. The RTI can act as an effective mechanism
for delivering high-density gas to the central gravitating source
\citep[e.g.,][]{KrumholzKMOC:09}. Thus, understanding of the RTI
around luminous gravitational sources such as proto-stellar cores or
BHs, is of critical importance for estimating their accretion rates,
which might enhance the final mass of stellar cores or the growth rate
of BHs. In particular, the accretion rate on to a BH is directly
related to the energy it emits but the enhanced luminosity may also
suppress the gas supply to the BH. Therefore, the onset of RTI could
be important for the mechanism of self-regulation of BH growth.

In this paper, we perform radiation-hydrodynamic simulations to
examine the special case of RTI in the context of radiation-regulated
accretion on to BHs. The main goal of this study is to test whether
the RTI plays an important role in supplying high-density gas to
the central BHs and to identify the physical conditions for the
stability of the I-front. In Section~\ref{sec:motivation}, we discuss
the phenomenology related to RTI found in our previous studies while we
explain the relevant physical scales and the numerical simulations
in Section~\ref{sec:method}. In Section~\ref{sec:results}, we show the
numerical results, and finally, we summarize and discuss the results 
in Section~\ref{sec:summary}.

\begin{figure*}
\includegraphics[width=180mm]{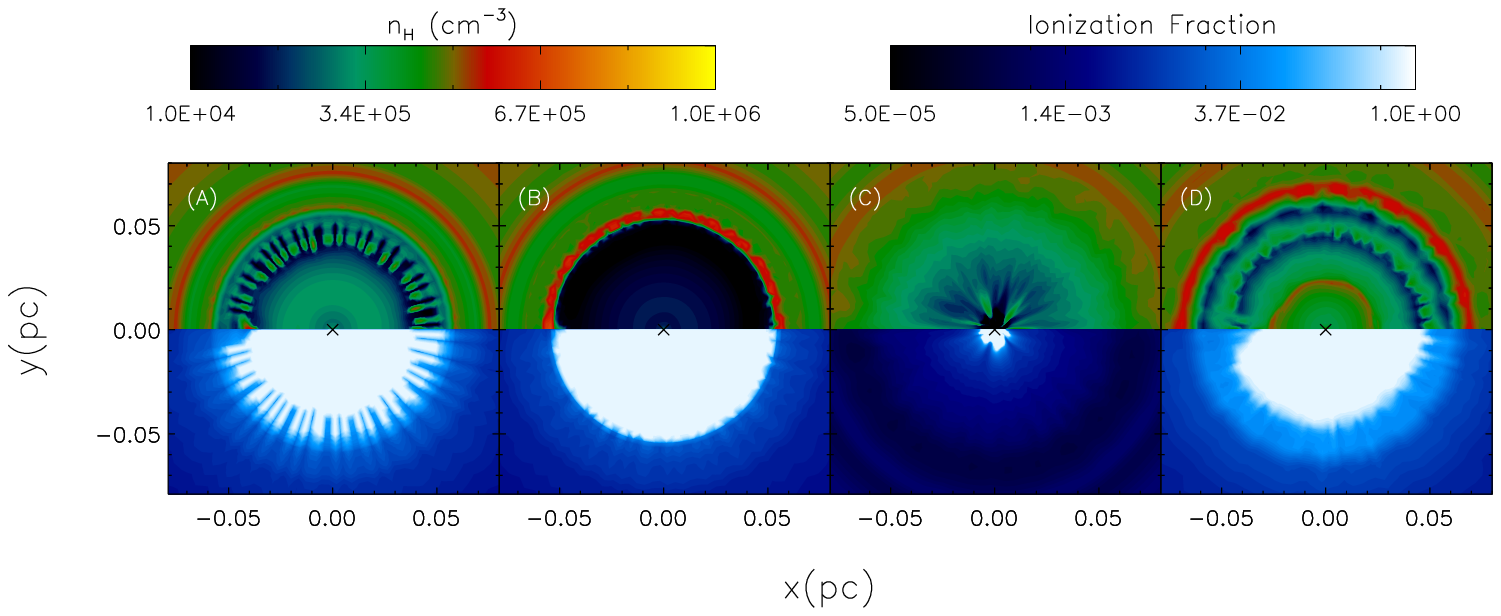}
\caption{Snapshots from our previous 2D radiation-hydrodynamic simulation
of radiation-regulated accretion on to a BH \citep{ParkR:11} of mass
$M_{\rm bh}=100~$M$_{\sun}$ surrounded by gas with density $\nH=\nfive$,
and temperature $T_\infty=10^4$~K. RTI at the I-front develops transiently
(A, C and D) when the BH luminosity increases mildly (A) or bursts (D),
or at the collapsing moment of the I-front (C). However, the RTI observed
throughout the simulations in various phase of the oscillation becomes
suppressed as shown in (B) maintaining the spherical symmetry of the
ionized \hii~region. The overall underlying hydrodynamic structure of the
\str~sphere is ideal for the growth of the RTI since a low-density gas
is surrounded by a high-density gas with an acceleration due to gravity;
 however, the radiation from the BHs suppresses the growth
of the RTI. The condition for the transitory growth of the RTI matches
well with the Equation~(\ref{eq:tau_rt}) implying that RTI can grow
temporarily when the density contrast between two media increases or the
effective acceleration on the I-front increases (i.e. when the I-front
expands outwards against the gravitational acceleration as shown in A
and D).}
\label{fig:motivation}
\end{figure*}

\begin{figure*}
\includegraphics[width=110mm]{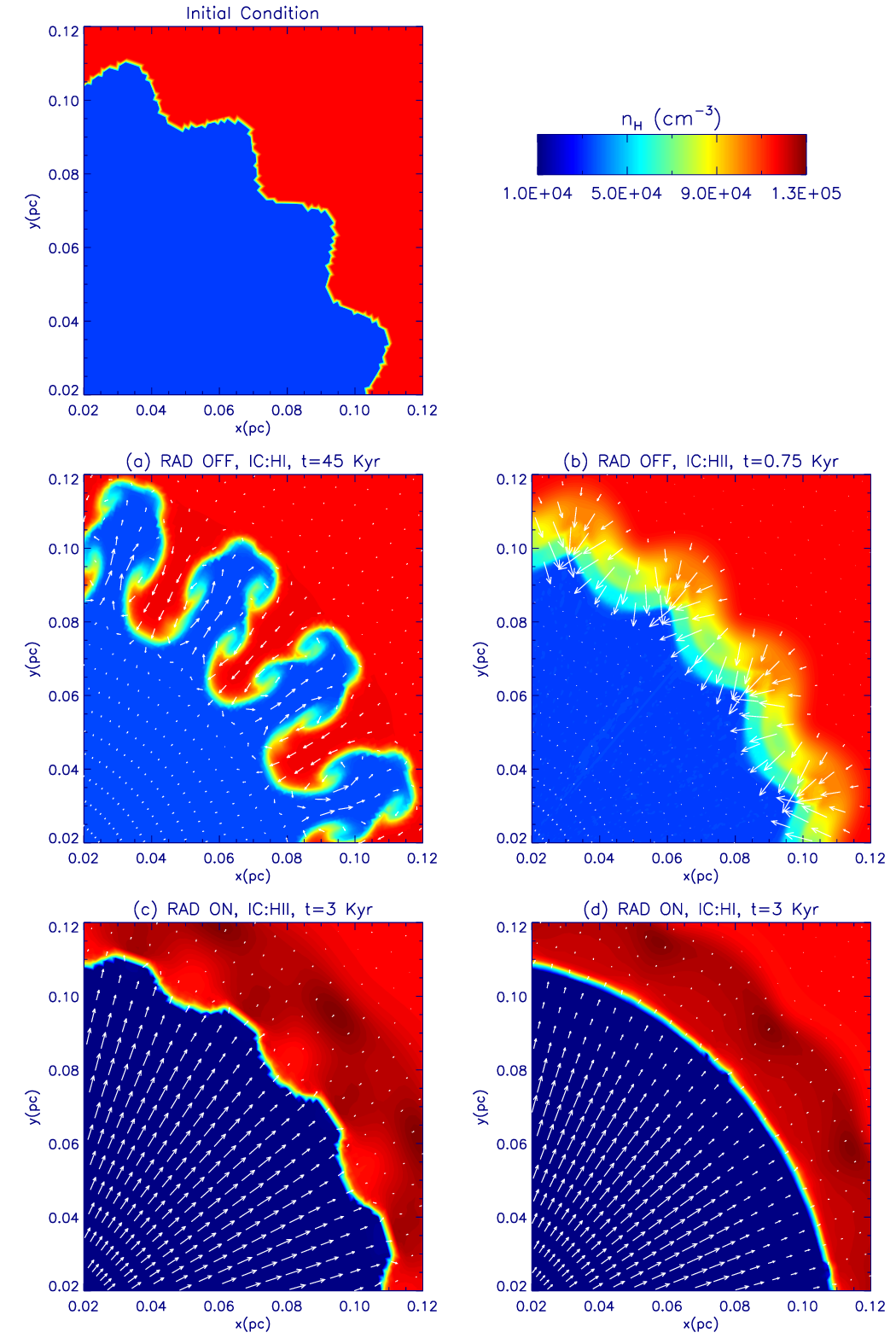}
\caption{Growth and suppression of RTI in various condition for a BH
of mass $M_{\rm bh}=100$~M$_{\sun}$ surrounded by gas with a density
$\nH=\nfive$. Initial temperature of the \str~sphere is set as $T_{\rm
in}=6\times 10^4~$K and the density is set as $n_{\rm H, in}
= T_\infty/T_{\rm in} \nH$ so that the IC is in thermal
equilibrium condition Top: density profile of the IC for
all simulations (a)--(d). Panel (a): snapshot of the density profile
taken at $t=45~$Kyr for the run K20RTI (no radiation from the
BH). Panel (b): density snapshot taken at $t=0.75~$Kyr for the run
K20RTIrec (low-density
region is ionized at the IC and the radiation from the BH remains turned off
during the simulation. Only the effect of recombination of ionized gas
is considered in this simulation).  
Panel (c): density snapshot taken at $t=3~$Kyr for the run K20HII
(radiation from the BH is on and the low-density region is ionized
at the IC).  Panel (d): density snapshot taken at $t=3~$Kyr for the
run K20HI (radiation from the BH is on and the low-density region
is neutral at the IC). Simulations become unstable to RTI when there is
no radiation from the BH shown in (a) and (b) while the radiation from
the BH suppresses the growth of RTI on the I-front effectively shown in
(c) and (d). Photoionization and the thermal pressure gradient
inside \hii~region effectively suppress the growth of the RTI. The recombination also dampens the initial perturbation effectively as shown in (b) and (d).} 
\label{fig:rad_norad}
\end{figure*}


\section{Motivation}
\label{sec:motivation}
UV and X-ray photons emitted from a BH accreting from a neutral
medium, create a hot and ionized bubble of gas (i.e., a \str sphere)
around the BH. A {\it D}-type I-front separates the ionized gas from the
neutral higher density gas. In \citet{ParkR:11,ParkR:12,ParkR:13}, we
explored the role of the \str sphere in regulating the gas supply from
large scales to the BH. We found a periodic variation of accretion
luminosity and thus the size of the \str~sphere \citep{MilosCB:09,
  Li:11} determined by the thermal structure of the gas inside the
\str sphere. \citet{ParkR:12} 
found that the average accretion
rate is typically 1 per cent of the Bondi accretion rate of the
neutral gas when it is radiation-regulated. In this regime, the average
accretion rate is comparable to the Bondi rate in the ionized gas
that is lower than the Eddington limit when $\nH M_{\rm bh} <
4\times10^8 ~{\rm cm}^{-3} {\rm M}_{\sun}$.
In general, the effect of photon momentum is found to be
negligible since the accretion rate shows oscillatory behaviour due to
self-regulation. The acceleration due to radiation pressure is negligible
compared to the gravity by the BH in the entire range of radius for most
time of the quasi-periodic oscillation. However, for the case of
momentum-driven I-front in the Eddington-limited regime, the photon
momentum can be effective in enhancing the RTI across the I-front since
radiation pressure increases the effective gravity at the I-front.

The \str sphere should be subject to RTI since lower density (i.e.,
hot and ionized bubble) gas supports higher density neutral gas
against the BH's gravity. Moreover, the feedback loop generates sharp
bursts of the BH luminosity, that result in a rapid increase of the
size of the \str~sphere. When the \str~sphere expands during the
luminosity burst, the RTI can grow on a shorter time-scale since the
effective acceleration in the frame of reference of the I-front
increases. However, our previous simulations \citep{ParkR:11,ParkR:12}
show that the I-front is remarkably stable to RTI, even for
perturbations with short wavelength that should grow on time-scales
shorter than the period between luminosity bursts when the \str~sphere
is stationary. In addition, 1D and 2D simulations produce the same
results in terms of accretion rate and period of the luminosity cycle.
We observe a transitory development of the RTI only during the
periodic luminosity bursts, however the growth of RTI fingers is
quickly suppressed and does not fuel the BH. Instead, we typically
observe a well-defined spherically symmetric low-density sphere that
varies in size as a function of time. Snapshots in
Fig.~\ref{fig:motivation} \citep[see also][]{ParkR:11} shows
the transient development and suppression of the RTI.  Panel (A) shows
the RTI fingers when the I-front expands as a result of an increase of
the BH luminosity, while panel (B) displays the suppression of the RTI
at the I-front and the recovered spherical symmetry of the \str
sphere. Front instabilities are typically observed just before and
after a luminosity burst. Panel (C) shows a snapshot at the time of
collapse of the \str~sphere that leads to a luminosity burst. This
happens when the BH luminosity reaches its minimum value in the cycle
and the gas density inside the \hii~region has its lowest value. Panel
(D) displays a snapshot at a time right after a luminosity burst, also
showing instabilities.

In our previous simulations we focused on the mechanisms of
self-regulation of gas accretion on to BHs. However, in those
simulations it is difficult to examine the suppression mechanism of the
RTI in detail because of the varying BH luminosity and size of the
\str~sphere. In this paper, we focus on the growth/suppression of the
RTI at I-front using controlled numerical experiments in which we keep
the BH luminosity, the size of the \str~sphere, the amplitude and
wavenumber of the perturbations fixed.   
We assume idealized spherically symmetric accretion on to stationary BHs
from non-magnetized neutral gas with zero angular momentum. In particular,
we investigate the RTI of I-front on the order of the Bondi radius scale
where the gravity becomes dominant over the thermal energy of the neutral
gas, and thus the RTI is effectively influenced. The present study,
although idealized, illustrates the stabilizing effect of opacity to
ionizing radiation to the development of RTI at I-fronts. The
simplifying assumptions of spherical symmetry and small gas angular
momentum are rather realistic at the scale of the Bondi radius
in many astrophysical applications (galactic winds or active galactic
nucleus winds); however, they are not critical to the development
of the instability.

\begin{table*}
 \centering
  \caption{Simulation parameters.}
  \begin{tabular}{ccccccccccc}
  \hline
	ID	&$M_{\rm bh}({\rm M}_{\sun})$	&$\frac{L}{L_{\rm Edd}}$ 	&$\nH (\rm cm^{-3})$ 		&$T_\infty$(K)
	&$\frac{\nH}{n_{\rm H, in}}$ &$R_{\rm s,0} {\rm (pc)}$ &$r_{\rm B}$ &$k$ &$\rm {IC} $ & $\delta R_{\rm s,0}/ R_{\rm s,0}$   \\
 \hline
	K20RTI    &$1\times 10^2$	&$0.0$		    &$10^5$  &$10^4$	& 6.0   &0.0	&0.005 &20 & \hi & $0.05$ \\
	K20RTIrec &$1\times 10^2$  	&$0.0$		    &$10^5$  &$10^4$	& 6.0   &0.0 	&0.005 &20 & \hii & $0.05$\\
	K20HI    &$1\times 10^2$	&$10^{-1}$	    &$10^5$  &$10^4$	& 6.0   &0.11	&0.005 &20 & \hi & $0.05$\\
	K20, K40, K80 &$1\times 10^2$	&$10^{-1}$	    &$10^5$  &$10^4$	& 6.0   &0.11	&0.005 &20, 40, 80 & \hii & $0.05$\\
	RANE1--RANE5  &$1\times 10^2$ &$10^{-1}$--$10^{-5}$ &$10^5$  &$10^4$	& 6.0   &0.11--0.005 &0.005 &496 &\hi & $0.1,~0.2$\\
	T4A      &$2\times 10^5$	&$5.0\times10^{-5}$ &$10$   &$10^4$	& 6.0   &33 	&9     &20 &\hii & $0.1$\\
	T4B      &$8\times 10^5$        &$1.25\times10^{-5}$ &$10$  &$10^4$	& 6.0   &33 	&35    &20 &\hii & $0.1$\\
	T3A      &$2\times 10^5$	&$5.0\times10^{-5}$ &$10$   &$10^3$	& 60.0  &130 	&90    &20 &\hii & $0.1$\\
	T3B      &$5\times 10^5$	&$2.0\times10^{-5}$ &$10$   &$10^3$	& 60.0  &130 	&230   &20 &\hii & $0.1$\\
	T3ABURST &$2\times 10^5$	&$5.0\times10^{-5}$ &$10$   &$10^3$	& 60.0  &130 	&90    &12 &\hi & $0.1$\\
\hline
\end{tabular}
\label{table:simulations}
\end{table*}

\section{Methodology} 
In this section, we explain the basic equations and typical scales related to
accretion physics, BH radiation, and RTI. We also describe the set-up
of our numerical simulations.

\label{sec:method}
\subsection{\str~Sphere in BH Systems}

The Bondi equations describe spherically symmetric accretion on to a
point source \citep{BondiH:44,Bondi:52}. The Bondi radius is the
distance from the point source where the BH's gravitational potential
roughly equals the thermal energy of the surrounding gas. It is defined as
$r_{\rm B} \equiv GM_{\rm bh}/c_{\rm s, \infty}^2$, where $M_{\rm bh}$ is
the BH mass and $c_{\rm s,\infty}$ is the sound speed of the
neighbouring gas. The Eddington luminosity $L_{\rm Edd}$ refers to the
maximum luminosity that a BH can achieve assuming that the radiation
pressure due to Compton scattering of photons on free electrons balances the BH's
gravity:
\begin{equation}
L_{\rm Edd} = \frac{4{\upi} GM_{\rm
bh}m_{\rm p}c}{\sigma_{\rm T}} \simeq 3.3\times
10^4 M_{\rm bh, \sun} {\rm L}_{\sun}.
\end{equation}
Here, $\sigma_{\rm T}$ is the Thomson scattering cross-section and $m_{\rm p}$
is the proton mass. The dimensionless BH luminosity $l$ is defined
as $l \equiv L/L_{\rm Edd}$. 
The number of ionizing photons $N_{\rm ion}$ emitted by the BH per unit time is 
\begin{equation}
N_{\rm ion} = \frac{L}{m\times 13.6~{\rm eV}} = 1.9 \times 10^{48} l M_{\rm bh, \sun} ~{\rm s^{-1}}, 
\end{equation} 
where $m=\alpha/(\alpha-1)$ depends on the slope $\alpha$ of the
power-law spectrum of the BH radiation (we take $\alpha=1.5$ in this
work). The size of the \str sphere around the BH can be calculated
from $N_{\rm ion}$ when the hydrogen number density, $n_{\rm H}$, is known:
$N_{\rm ion}=(4\pi/3) R_{\rm s}^3 n_{\rm e} n_{\rm H} \alpha_{\rm rec}$, where we assume
hydrogen recombination coefficient $\alpha_{\rm rec}=4\times 10^{-13}
(T_{\rm in}/10^4)^{-1/2}~$cm$^3$s$^{-1}$. Therefore, the size of the
\str sphere is
\begin{equation}
\ars = 2.5 \times 10^{-3} \left( \frac{ l M_{\rm bh, \sun} }{ n_{\rm H}^2 \alpha_{\rm rec}} \right)^{1/3}~{\rm pc}. 
\label{eq:ars}
\end{equation}
The assumption of thermal pressure equilibrium across the I-front is a
good approximation only if $\ars > r_{\rm B}$. Instead the I-front collapses
on to the BH if $\ars < r_{\rm B}$, or equivalently, when the thermal
energy of the gas outside the \str~sphere is smaller than the
gravitational potential energy at the I-front:
\begin{equation}
c_{\rm s, \infty}^2 < \frac{GM_{\rm bh}}{R_{\rm s}}. 
\end{equation}

\begin{figure}
\includegraphics[width=85mm]{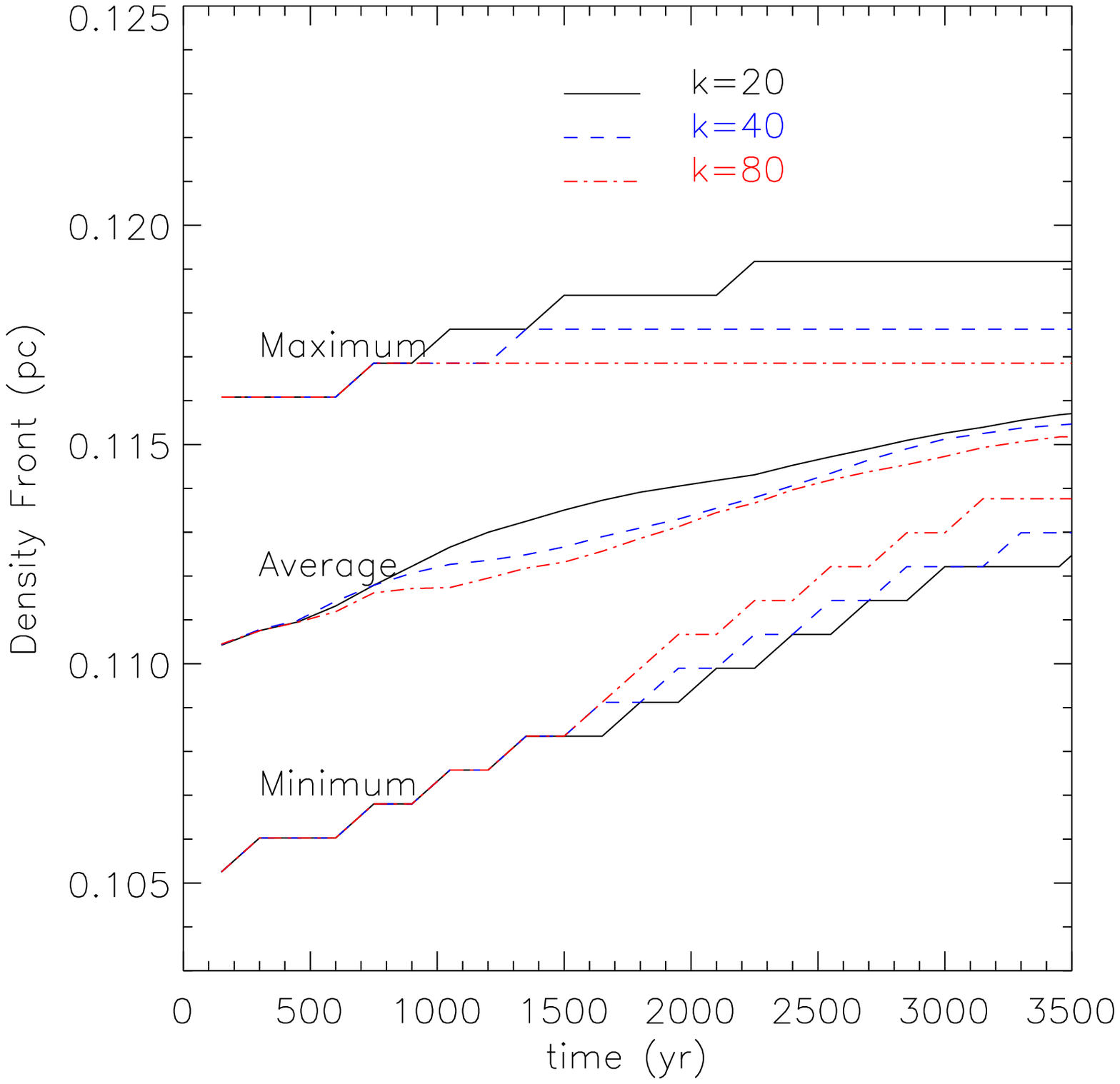}
\caption{Suppression of RTI for different wavenumbers $k=$20 (solid)
, 40 (dashed) and 80 (dot--dashed). From top to bottom, each group of
lines shows the maximum, average and the minimum loci of the I-fronts.
Note that the average locations increase as a function of time (by about
5 per cent at $t=3500~$yr) due to the decreased density inside the
\hii~region. The difference between the maximum and minimum locations
decreases for each simulation as a function of time for all values of
$k$. Simulations with bigger wavenumbers (i.e. shorter wavelengths)
reach the spherical symmetry on slightly shorter time-scales
with the initial perturbation suppressed by the radiation.}
\label{fig:dfront} 
\end{figure}

\begin{figure*}
\includegraphics[width=170mm]{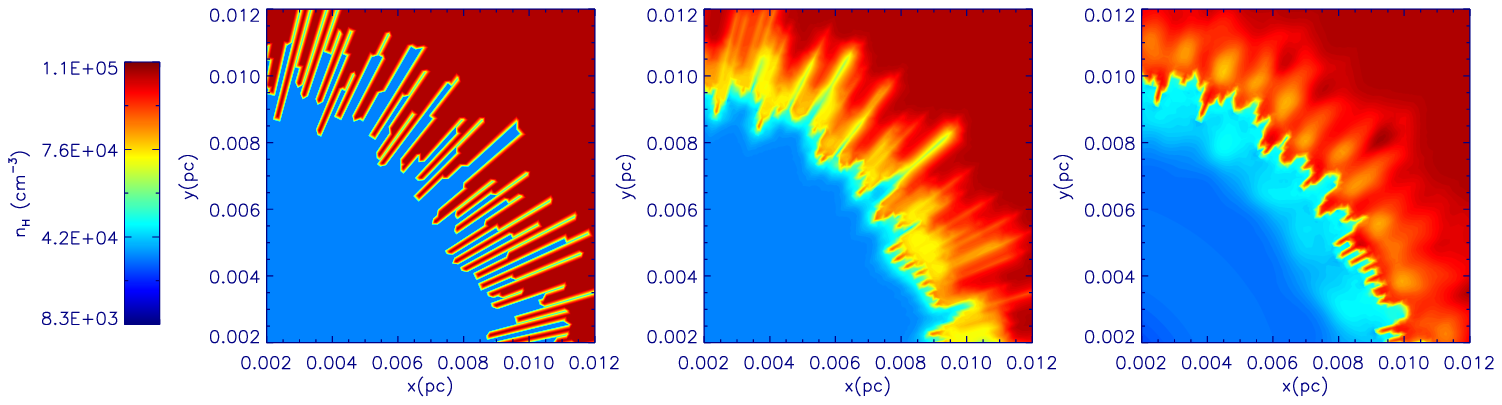}
\caption{Simulation {RANE3} for a BH mass $M_{\rm bh}=100$~M$_{\sun}$, gas density 
$\nH=\nfive$ and the luminosity in Eddington unit $l=0.001$
with a random perturbation of I-front with a maximum of 20 per cent of
the initial $R_{\rm s}$. The smallest wavelength of $\sim 0.0001~$pc set by the
simulation resolution does not grow, but becomes suppressed by the 
photoionization and recombination effect.}
\label{fig:random}
\end{figure*}

\begin{figure*}
\includegraphics[width=170mm]{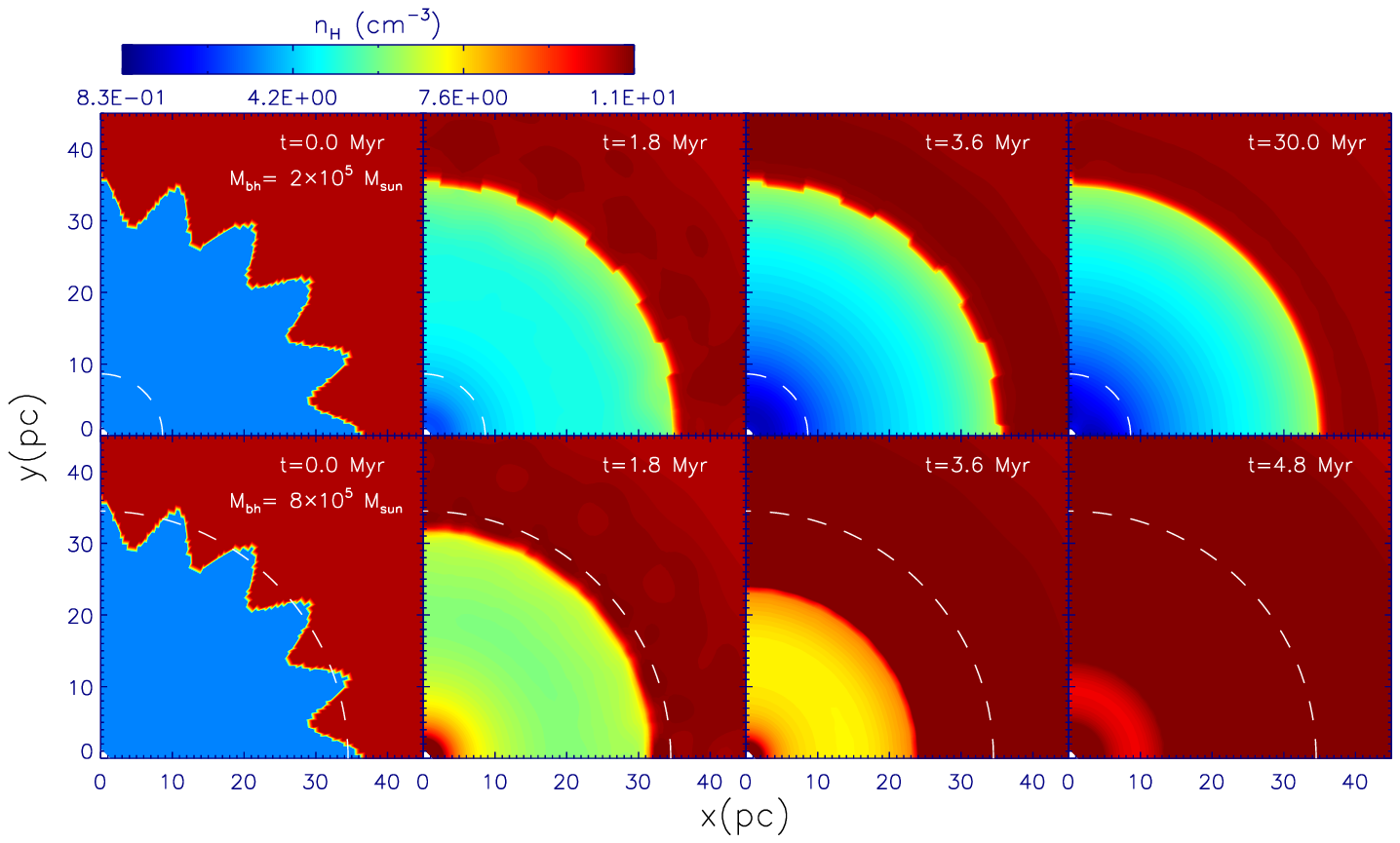}
\caption{Top: evolution of the \hii~region for the run T4A
with a BH mass $M_{\rm bh}=2\times 10^5~$M$_{\sun}$, gas temperature $T_\infty =
10^4$~K and density $\nH=\none$. Each panel displays the density profiles
at $t=0.0, 1.8, 3.6$~and $30.0$~Myr, respectively. Bottom: evolution of the run
T4B with the same initial density, temperature, and BH luminosity
as the top panels, but with a larger BH mass $M_{\rm bh}=8\times
10^5~$M$_{\sun}$. Snapshots are taken at $t=0.0, 1.8, 3.6$~and
$4.8$~Myr, respectively. Dashed lines show the Bondi radius $r_{\rm B}$ for
each BH mass $M_{\rm bh}$. The Bondi radius for the run T4A is located
within the \hii~region while the Bondi radius is comparable to the
size of \hii~region for the run T4B. When the Bondi radius is comparable
to or smaller than the size of the \hii~region, the entire \hii~region 
shrinks to the BH with the radiation trapped in the gas inflow.}  
\label{fig:t4}
\end{figure*}

\subsection{RTI at the I-front}

In the linear regime, the growth rate of the RTI at the interface
between high-density ($\rhoH$) and low-density ($\rhoL$) media
for a wave number $k$ ( $= 2\pi/\lambda$) is $\gamma = \sqrt{\mathcal{A}
g k}$, where the Atwood number is defined as $\mathcal{A} \equiv (\rhoH -
\rhoL)/(\rhoH+\rhoL)$. Therefore the growth time-scale of the RTI $\tau_{\rm
RT}$ for linear perturbations with wavelength $\lambda$ is
\begin{equation}
\tau_{\rm RTI} = \gamma^{-1} = \sqrt{\frac{\rhoH + \rhoL}
{\rhoH - \rhoL}\frac{\lambda}{2\upi g}} \sim \sqrt{\frac{\lambda}{2\upi g}}, 
\label{eq:tau_rt}
\end{equation}
where $g = GM_{\rm bh} \ars^{-2}$ is the gravitational acceleration at
the I-front and $\mathcal{A}$ is close to unity if $\rhoH \gg
\rhoL$. Assuming that the I-front is stationary (i.e. is not
accelerating: $L={\rm const}$, and is not collapsing: $\ars > r_{\rm B}$), the RTI
growth time-scale is
\begin{equation}
\tau_{\rm RTI} \sim \sqrt{\frac{\lambda}{2\upi g}} 
= \sqrt{\frac{\lambda \ars ^2}{2\upi GM_{\rm bh}}}  
> \sqrt{\frac{\lambda \ars}{2\upi c_{\rm s, \infty}^2}}. 
\label{eq:tau_rt_cs}
\end{equation}
It is well known that recombination in the ionized gas is a
stabilizing mechanism for I-fronts \citep{Mizuta:05}. This notion
suggests that the RTI can be stabilized by recombination only if
the RTI time-scale $\tau_{\rm RTI}$ is longer than the recombination
time-scale $\tau_{\rm
  rec} \sim 1/n_{\rm H} \alpha_{\rm rec}$. From this, we obtain
  that
perturbations with wavelength $\lambda>\lambda_{\rm crit}$ are
stabilized by recombination, where the critical wavelength is
\begin{equation}
\lambda_{\rm crit} \equiv \frac{2\upi G M_{\rm bh} \tau_{\rm rec}^2}{\ars^2}. 
\label{eq:lambda_crit}
\end{equation}
If we define a critical angular scale of the perturbations
$\theta_{\rm crit} \equiv \lambda_{\rm crit}/2\upi R_{\rm s}$, using
Equations~(\ref{eq:ars}) and (\ref{eq:lambda_crit}) we find that scales $\theta>\theta_{\rm crit}$ are stable, where 
\begin{equation}
\theta_{\rm crit} \sim 5\times 10^{-15}\frac{G}{\alpha_{\rm rec} l }  \sim \frac{2\times 10^{-9}}{l}.  
\label{eq:theta_crit}
\end{equation}
Thus, $l$ needs to be very small ($\sim 10^{-9}$) in order to
observe the growth of RTI on scales comparable to the \str~radius.
Given the limited resolution of finite grids of the
current study, the development of the RTI can be observed only when
$l$ is extremely low or the front is accelerating outwards, which makes
the critical scale $\theta_{\rm crit}$ increase. This is indeed what
has been observed in the simulations presented in \citet{ParkR:11,ParkR:12}.
If the effective gravitational acceleration at the I-front increases
due to the front acceleration, larger wavelength can become
unstable.  
The effective gravitational acceleration on the I-front
increases when the I-front propagates against the gravity moving away
from the BH. This occurs when the BH luminosity increases (see panels
A and C in Fig.~\ref{fig:motivation}). 
See \citet{WhalenN:08}
for details about the radiation hydrodynamic instabilities for
propagating I-fronts.


Although we found that if $l \le 10^{-9}$ the I-front can be unstable
to RTI, in this case we also expect that the I-front cannot
be in equilibrium against the gravitational attraction of the BH
because we are in the regime $R_{\rm s} < r_{\rm B}$. If we compare the RTI growth
time-scale $\tau_{\rm RTI} \sim \tau_{\rm rec}$ to the free-fall time-scale
$\tau_{\rm ff}=\ars^{3/2}(2GM_{\rm bh})^{-1/2}$ we find $\tau_{\rm
  RTI}/\tau_{\rm ff} \propto l^{-1}$ implying that when $l$ is small
the RTI growth is slower than the time it takes for the I-front to
collapse on to the BH.

\subsection{Radiation-hydrodynamic Simulations}
\label{sec:numerical}

We run a set of 2D radiation-hydrodynamic simulations to examine the
growth of the RTI at the I-front produced by radiation emitted by a
BH. We use a parallel version of the non-relativistic hydrodynamic
simulation code \textsc{zeus-mp} \citep{StoneN:92,Hayes:06} with our 1D
radiative transfer equation solver \citep{RicottiGS:01,WhalenN:06}.
We use an operator-split method switching between hydrodynamic and
radiative transfer steps. For every time step, we use the smallest between the
hydrodynamic and chemical time steps $ \Delta t = {\rm min} (\Delta
t_{\rm hydro}, \Delta t_{\rm chem})$. Our 1D radiative
transfer subroutine calculates photoionization (and thus the chemistry
of \hi, \hii, {\rm He~{\textsc i}}, {\rm He~{\textsc{ii}}}, {\rm
  He~{\textsc{iii}}} and $e^-$), photoheating, and cooling. We assume
that UV and X-ray photons are emitted from the BH centred at the
origin of the spherical coordinate following a power-law spectrum
($F_\nu \propto \nu^{-\alpha}$) with a spectral index $\alpha=1.5$ in
the range of $13.6$~eV--$100~$keV in 50 logarithmically spaced
frequency bins.  Radiative transfer equations are solved for each
angular direction ($\theta$) every time step. Radiation pressure on
electron and neutral hydrogen \citep{ParkR:12} is not considered in
the current work since simulations that included radiation pressure
did not show significant differences from simulations that did not
\citep{ParkR:11}.     
\subsubsection{Setting Up Initial Conditions } 

We adopt a constant luminosity, $l$, for a fixed BH mass, $M_{\rm
  bh}$. The temperature ($T_\infty=10^4~$or $10^3~$K) and the density
$\nH$ of the surrounding gas are also constant in our initial
conditions (ICs). From a given set of parameters, we calculate the
\str~radius using Equation~(\ref{eq:ars}) and a lower density sphere
with radius equal to the \str~radius is placed centred on the BH in
the ICs set-up. The gas is initially assumed to be in pressure
equilibrium. Thus, given the typical temperature $T_{\rm in}= 6\times
10^4~$K in the lower density ionized sphere \citep{ParkR:12}, the gas
number density inside the \str~radius is $n_{\rm H,
  in}=T_{\infty}/T_{\rm in}~n_{\rm H, \infty}$. We perturb the
spherical I-front by applying sinusoidal perturbations as $R_{\rm s}(\theta)
= R_{\rm s,0} (1+\delta R_{\rm s,0}/R_{\rm s,0}~{\rm sin}(k \theta ))$, where $k$
is the wavenumber. The amplitude of the perturbations $\delta
R_{\rm s,0}$ is a free parameter that we choose between $5$, $10$ and
$20$ per cent of $R_{\rm s,0}$. For simulations K20E1--K20E5, we introduce
random perturbations at the I-front for each radial direction resolved
in the simulation.  The Bondi radius is constant for a given
temperature $T_\infty$ and BH mass $M_{\rm bh}$, while we vary the BH
luminosity and thus the size of the \hii~region for a range of
luminosities $l$ between $0.1$ and $10^{-5}$ (see
Table~\ref{table:simulations}). In some simulations we fix the
absolute BH luminosity $L$ and vary the BH mass in order to explore
the effect of changing the ratio $r_{\rm B}/R_{\rm s}$.

We perform 2D simulations in polar
coordinate system ($r$, $\theta$) with BHs centred at the origin
assuming it is axisymmetric around $\theta=0$.
Evenly spaced $128$--$256$ grids
in the radial ($r$) and polar angle ($\theta$) directions are used
to obtain higher resolution at the I-front. In the
radial direction, we set the box size of simulations twice the
initial \str radius $R_{\rm s,0}$ with flow-in boundary conditions and
inner boundary between $0.02$ and $0.2$ of $R_{\rm s,0}$ with flow-out
boundary conditions. The angular direction ($\theta$) extends from
$0$ to $0.5\upi$ and reflective boundary conditions are applied.

\section{Results}
\label{sec:results}
\subsection{Suppression of RTI by Photoionization and Recombination}

Simulations shown in panels (a)--(d) of Fig.~\ref{fig:rad_norad}
have the same IC (shown in the top panels): gas density
($\nH=\nfive$), temperature ($T_\infty =10^4~$K) and perturbation
wavelength ($k=20$).  Each simulation differs for the presence/absence
of a ionizing radiation field emitted by the BH and the initial
ionization fraction inside the low-density bubble as summarized at the
top of each panel (see Table~\ref{table:simulations}).

Panels (a) and (b) in Fig.~\ref{fig:rad_norad} show snapshots for
simulations without radiation from the central BH ({K20RTI} and
{K20RTIrec} in Table~\ref{table:simulations}). The low-density
region in the simulation {K20RTI} is neutral (\hi) while in simulation
{K20RTIrec} the gas is ionized (\hii). Snapshot (a) shows an
exponential growth of RTI when the simulation evolves to $t=45000$
yr. The high- and low-density media are initially in thermal
equilibrium, but they start to mix with each other when the
high-density gas accelerates on to the low-density region while the
low-density fingers accelerate into the high-density region due
to buoyancy. Panel (b) shows the snapshot taken at $t=7500$~yr for
the simulation {K20RTIrec} which shows the effect of the
recombinations in the initially ionized gas. Initial perturbations at
the I-front are damped on a shorter time-scale compared to the typical
RTI growth time-scale (panel a). As the simulation proceeds, the
ionized gas recombines and the conditions become similar to the
typical RTI condition as in panel (a).

\begin{figure}
\includegraphics[width=80mm]{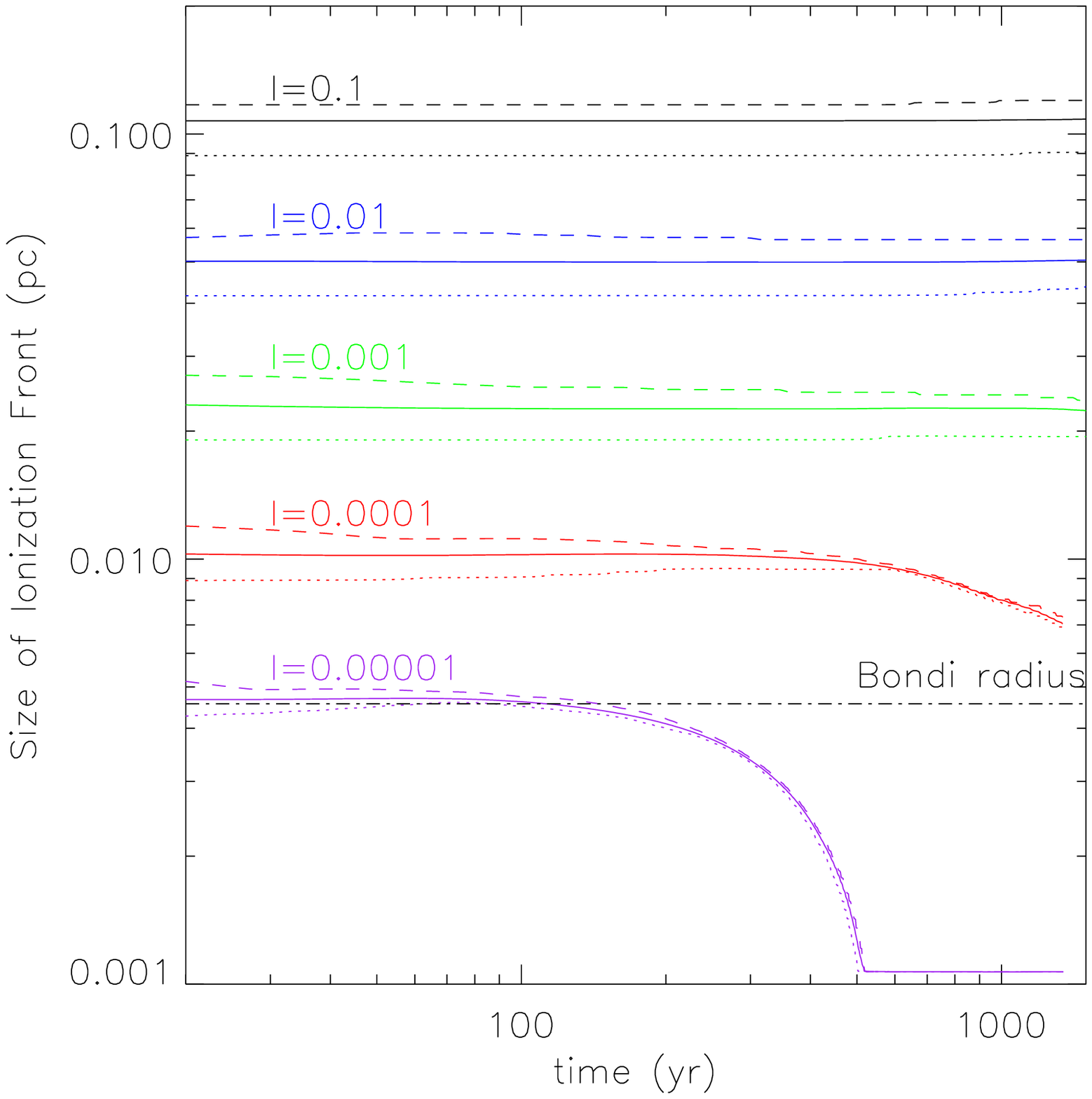}
\caption{Stability of I-fronts as a function of the Eddington ratio 
$l=L/L_{\rm Edd}$ for a BH with mass $M_{\rm bh}=100~$M$_{\sun}$ and gas
density $\nH=\nfive$. When $l > 0.001$, the size of the \hii~region
$\ars$ is larger than the Bondi radius making the \str~sphere stable
while the \hii~region shrinks to the BH when $l \le
10^{-5}$, where the $\ars$ becomes comparable to or smaller than the
Bondi radius. This transition occurs because the sound crossing time-scale
$\tau_{\rm cross}$ is shorter than the recombination time-scale
$\tau_{\rm rec}$. The amplitude of the initial perturbation
decreases as a function of time not allowing the growth of RTI since 
the $\tau_{\rm cross}$ is shorter than the $\tau_{\rm RT}$.}
\label{fig:edd_growth} 
\end{figure}

However, when photoionization is turned on (panels c and d in
Fig.~\ref{fig:rad_norad}), the simulations show that the growth of
RTI fingers is suppressed at the I-front. The bottom panels in
Fig.~\ref{fig:rad_norad} show the simulations {K20HI} and {K20HII}
with the radiation from the central BHs turned on, with the different
initial ionization fraction of the low-density bubble (neutral gas in
panel c and ionized gas in panel d). Both panels (c) and (d) show
that the RTI fingers are suppressed. Dense fingers extended into the
low-density region at the ICs are suppressed by the photoionization
while the low-density fingers (in the dense medium) are also suppressed
due to recombination. The I-front shown in panel (d) displays a
smoother spherical shape compared to the one in panel (c). This is due
to the effect of recombination since the I-front ionizes the inner
region first and recombination dampen the perturbation before the
I-front reaches the density discontinuity. The vector field of the
bottom panels of Fig.~\ref{fig:rad_norad} indicates that outflow
inside the \hii~region also contributes in suppressing the inward
motion of the dense RTI fingers as shown in panel (a). The outflow inside the \hii~region occurs due to the thermal
pressure gradient as discussed in \citet{ParkR:11,ParkR:12} and
serves as an important gas depletion mechanism inside the \hii~region.
Thermal pressure gradient inside the {\hii}~region dominates the
gravitational potential in the outer part of the \str sphere causing
the outflow seen in the bottom panels of Fig.~\ref{fig:rad_norad}.
Note that simulations (c) and (d) do not reach a steady state due
to the gas depletion inside the
\hii~region; however, the growth of RTI is not observed.

\subsection{Suppression of the RTI as a function of wavenumber}
 
In the linear regime of RTI growth, small-scale perturbations grow faster than
large scales (see equation~\ref{eq:tau_rt}). Here, we examine
how different wavelengths of RTI are affected by radiation.

Fig.~\ref{fig:dfront} shows the evolution of the maximum, average
and minimum location of I-fronts in simulations {K20}, {K40} and {K80}
in Table~\ref{table:simulations} that differ for wavenumber of the
initial perturbations: $k=20$ (solid lines), $40$ (dashed lines ) and
$80$ (dot--dashed lines). The average size of \str~sphere in all
simulations increases as a function of time due to the decreasing
density inside the \hii~region (the increase is about $5$ per cent of
the initial $R_{\rm s}$ at $t=3500$~yr). Relative locations of
minimum/maximum density fronts converge asymptotically to the averages
implying that the initial amplitude of the perturbations is
damped. Note that the minimum radii of the I-fronts (RTI fingers)
increase monotonically while the maximum loci do not increase much.
The simulations also show that the amplitude of perturbations with
larger wavenumbers (i.e. smaller wavelength) are suppressed on a
shorter time-scales.

In Fig.~\ref{fig:random}, we show a simulation with random
perturbations of the I-front in the ICs. The wavelength of the
perturbations produced this way is limited by the angular resolution
of the simulation.  In this simulation, the maximum amplitude of the
perturbation is $20$ per cent of the size of \hii~region and the
resolution in polar angle direction is $\ars \Delta \theta =
0.01\times 0.5\upi / 128 = 0.0001~$pc. These simulations do not show
any growth of the RTI even at the smallest scales permitted by the
resolution. Qualitatively, the evolution is the same as in simulations
with sinusoidal perturbations. With
$l=0.1$ the critical
angular scale from Equation~(\ref{eq:theta_crit}) is $\theta_{\rm
  crit} \sim 10^{-11}$ radian implying that RTI on smallest scales
that are not resolved in our simulations might not grow.

\begin{figure}
\includegraphics[width=80mm]{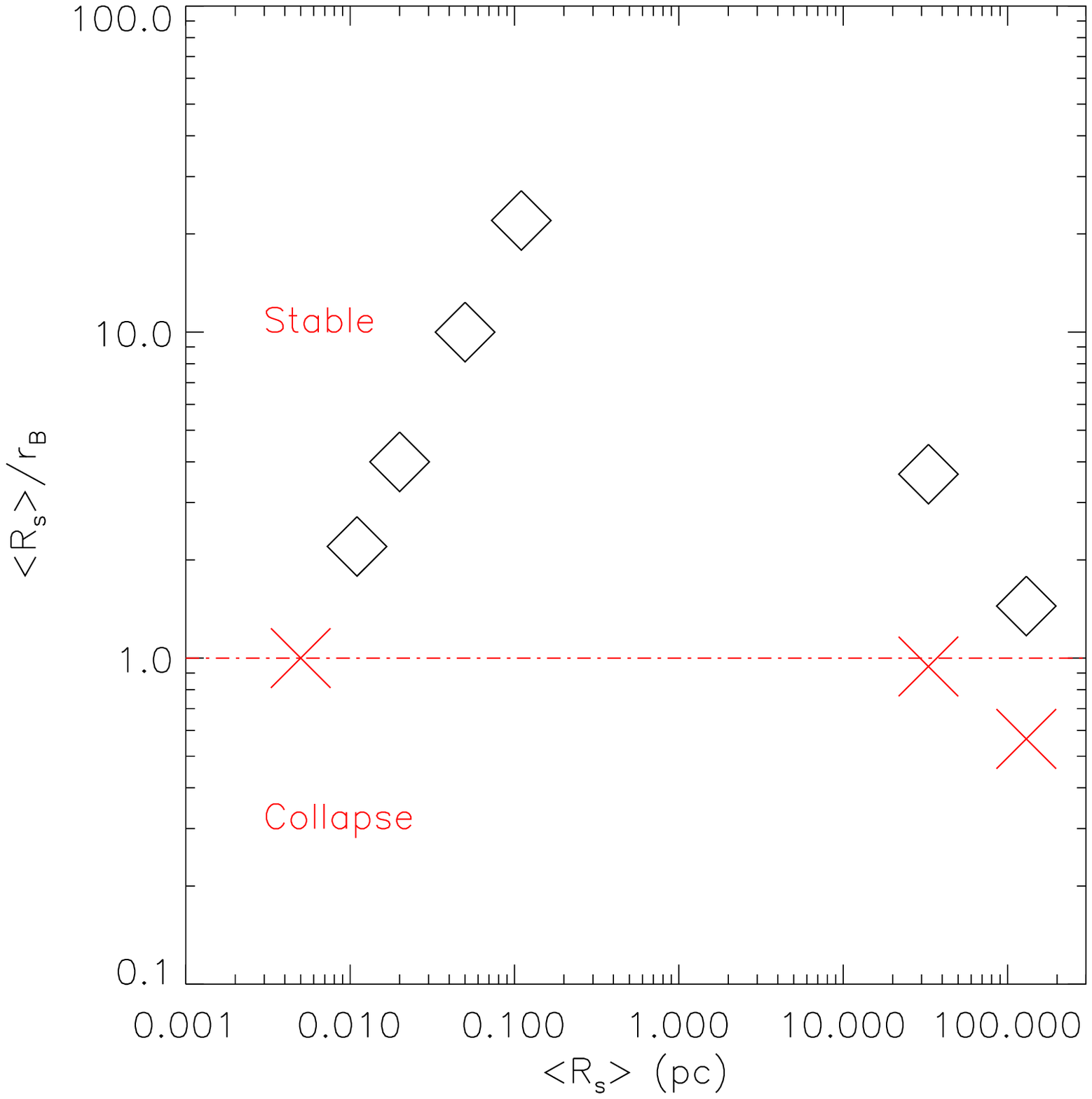}
\caption{Stability of the I-front for various ratio between $R_{\rm s}$ and
$r_{\rm B}$. Diamonds indicates the simulations (RANE1--RANE4, T4A and T3A)
showing stable I-fronts while crosses show the unstable simulations
(RANE5, T4B and T3B) with collapsing \str~spheres. When $\ars/r_{\rm B} \le
1$, \str~sphere shrinks to the BH not being able to maintain the initial
size of the \str~sphere. }
\label{fig:rs_rb} 
\end{figure}

\subsection{Stability of I-fronts}

In this section, we examine the stability of I-fronts varying the
gravitational acceleration at the I-front.  We vary the BH mass
keeping the BH luminosity and $R_{\rm s}$ fixed. Hence, the luminosity in
Eddington units $l$ varies.

Fig.~\ref{fig:t4} shows the evolution of  simulations with stable
(upper panels) and unstable (lower panels) I-fronts. The stability of
the I-front depends on the ratio $R_{\rm s}/r_{rm B}$ ($R_{\rm s}$ and $r_{\rm B}$ are shown
as dashed lines). In this set of simulations (T4A/T4B and T3A/T3B),
the BH luminosity is fixed and we set the BH mass $M_{\rm bh}=2\times
10^5~$M$_{\sun}$ in simulation T4A and $M_{\rm bh}=8\times
10^5~$M$_{\sun}$ in simulation T4B. The Bondi radius $r_{\rm B}$ is larger
than $\ars$ in the run {T4A} while $r_{\rm B}<\ars$ in run T4B. The
simulation sets T3A and T3B are analogous, but the temperature of the
ambient gas is $T_\infty =10^3~$K (instead of the fiducial value
$T_\infty =10^4~$K). The temperature of the gas inside the \str
sphere is $T_{\rm in}=6\times 10^4$~K; thus, the density inside the
\str~sphere is 60 times lower the ambient density (see
Table~\ref{table:simulations}).

Fig.~\ref{fig:edd_growth} shows the stability of the I-front as a
function of the luminosity $l$. When $l$ is larger than $10^{-3}$, the
I-fronts are stable since the $\ars$ is larger than the Bondi radius,
while the I-fronts collapse towards the BH for $l \le 10^{-5}$ when
$\ars \sim r_{\rm B}$. Note that the initial perturbations are suppressed
even when the I-front collapses.

Fig.~\ref{fig:rs_rb} shows the stability of the I-fronts for all the
simulations listed in Table~\ref{table:simulations} showing that the
ratio $\ars/r_{\rm B}$ determines the stability of the I-front. Simulations
RANE5, T4B and T3B in Table~\ref{table:simulations} have $\ars/r_{\rm B}
\le 1 $ and indeed their I-fronts collapse on to the BHs while the
I-fronts are stable in all the other simulations that have $\ars/r_{\rm B} >
1$. In all simulations the initial perturbations at the I-front are
suppressed creating a spherically symmetric \hii~region, however, when
$\ars/r_{\rm B} \le 1$ the ionized region collapses on to the central BH with
the radiation from the BH trapped by the inflowing gas.

\begin{figure*}
\includegraphics[width=170mm]{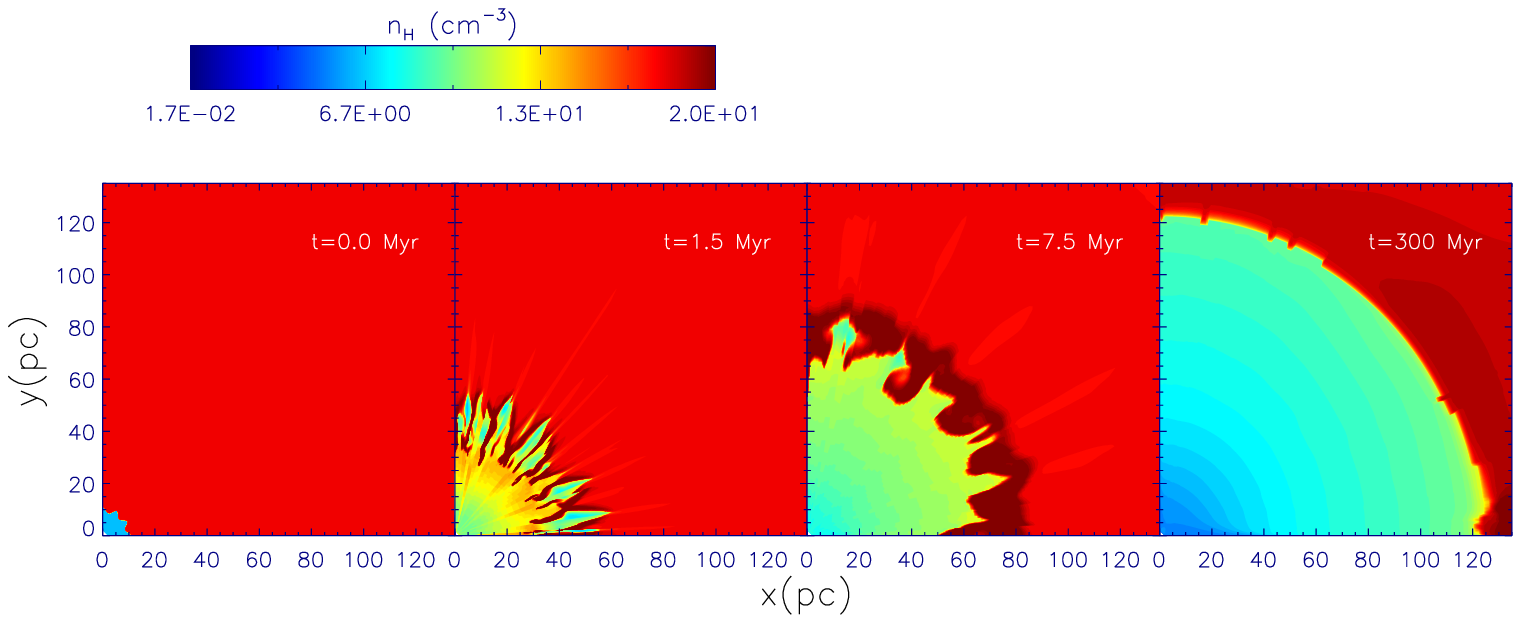}
\caption{Snapshots for a simulation T3ABURST with a luminosity burst.
Most of the parameters at the IC is identical to the simulation T3A which
shows the suppressed RTI, but the initial size of $R_s$ for this run is
10 times smaller. Since the BH luminosity is large enough to create a
10 times larger \hii~region, the \hii~bubble expands until it finds
an equilibrium (at $R_s \sim 130~$pc). Due to the expansion of the
\hii~region, RTI is observed at the initial phase of the simulation,
however RTI becomes suppressed as the I-front reaches the equilibrium
radius which is set by the BH luminosity.}
\label{fig:burst}
\end{figure*}

\subsection{RTI with a burst of luminosity} 

Finally, we simulate the case when the I-front propagates outward
against the gravity, increasing the value of the effective
acceleration in the frame of reference of the I-front.
Fig.~\ref{fig:burst} shows the evolution of the \hii~region which
expands following a luminosity burst. The initial luminosity is the
same as in simulation {T3A} corresponding to a \str~sphere with $\ars
= 130~$pc. But here we decrease the size of the low-density region by
a factor of 10 in the ICs.  The I-front expands since the equilibrium
size of \hii~region for the given BH luminosity is larger than the
radius of the contact discontinuity in the ICs. The development of the
RTI is observed at the beginning of the simulation during the outward
propagation of the I-front. However, the RTI is suppressed as the
I-front reaches the \str~radius set by the BH luminosity. We thus
conclude that stationary I-fronts produced by BHs accreting from a
neutral medium are stable to the development of RTI. Only an
accelerating I-front produced by an accretion and luminosity burst
develops RTI.

\section{Summary and Discussion}
\label{sec:summary}

In this paper, we examine the growth and suppression of RTI at the
I-front around BHs. This is a follow-up study on our previous works on
radiation-regulated accretion on to BHs \citep{ParkR:11,ParkR:12}. We
use numerical techniques that couple radiation to hydrodynamics and
focus on the condition of RTI development at the I-front. In general,
our radiation-hydrodynamic simulations do not show growth of RTI in the
presence of the ionizing radiation from BHs, which agrees 
with the perturbation analysis of the linear growth of RTI \citep{Ricotti:13}. The
following is a summary of the main findings: 
\begin{itemize}
\item For non-accelerating I-fronts, the RTI is stabilized on
all scales resolved in our simulations. 
\item Our calculations show
that the wavelength of the unstable modes is inversely proportional to
the luminosity $l$ in Eddington units, but when $l$ is small the I-front collapses on to the BH
for because $R_{\rm s} < r_{\rm B}$, preventing the development of the RTI in all realistic cases.
\item Accelerating I-fronts, produced during a luminosity burst, become
  unstable to RTI but are quickly stabilized by the radiation when the
  burst ends.
\end{itemize}

The findings in this study are relevant for understanding the role of
radiation in suppressing the RTI at the I-front around luminous
gravitating sources such as BHs or proto-stars. For quasi-spherically
symmetric systems, the RTI is not an efficient mechanism for
delivering gas from large scales to the central gravitating
source. For systems with high angular momentum, the spherical symmetry
assumed in this study is not realistic and the RTI might play an
important role in particular directions where the radiation flux is
reduced due to enhanced gas column density
\citep[e.g.,][]{KrumholzKMOC:09}.  Due to the limited degrees of
freedom in our 2D simulations, more realistic 3D simulations might
display slightly different results. Nonetheless, the qualitative
description of the role of photoionization on RTI presented in this
work should still be valid.

\section*{Acknowledgements}
This research is supported by the Urania E. Stott Fellowship of The
Pittsburgh Foundation. MR's research is supported by NASA grant
NNX10AH10G, NSF CMMI1125285 and ILP LABEX (under reference
ANR-10-LABX-63) supported by French state funds managed by the ANR
within the Investissements d'Avenir programme under reference
ANR-11-IDEX-0004-02. TDM acknowledges the National Science Foundation, 
NSF Petapps, OCI-0749212 and NSF AST-1009781 for support.
KP thanks James Drake for the motivation of this paper, and the
anonymous referee for useful comments. The numerical simulations
in this paper were performed using the computer cluster (``yorp")
of the Center for Theory and Computation, the Department of Astronomy
at the University of Maryland at College Park.



\label{lastpage}

\end{document}